\documentclass{osa-article}
\journal{osajournal}
\articletype{Research Article}

\newcommand{\vvv}[1]{\mathbf{#1}}
\newcommand{\xhatv}{\mathbf{\hat{x}}}

\begin{document}

\title{Spatiotemporal Characterization of Nonlinear Interactions between Selectively Excited Radially Symmetric Modes of a Few-Mode Fiber}

\author{Sai Kanth Dacha\authormark{1,2,*}, Thomas E. Murphy\authormark{1,3}}

\address{\authormark{1}Institute for Research in Electronics and Applied Physics, University of Maryland, College Park, MD 20740, USA}
\address{\authormark{2}Department of Physics, University of Maryland, College Park, MD 20740, USA}
\address{\authormark{3}Department of Electrical and Computer Engineering, University of Maryland, College Park, MD 20740, USA}

\email{\authormark{*}sdacha@umd.edu} 

\begin{abstract}
Nonlinear propagation of signals in single-mode fiber is well understood, and is typically observed by measuring the temporal profile or optical spectrum of an emerging signal.  In multimode fibers, the nonlinearity has both a spatial and a temporal element, and a complete investigation of the interactions between propagating modes requires resolving the output in both space and time.  We report here spatiotemporal measurements of two nonlinear interacting LP$_{0m}$ modes of a step-index few-mode fiber (FMF).  We describe a method to selectively excite two propagating modes through the use of a phase-mask directly patterned on the entrance face of the fiber.  The output is resolved by raster-scanning a near-field tapered single-mode optical fiber probe that is connected to a high-speed detector.  The results show that in the presence of nonlinearity, the output exhibits a spatiotemporal character that cannot be adequately characterized by a camera image or pulse shape alone.
\end{abstract}

\section{Introduction}

The rapid growth in demand for long-haul fiber-optic communication systems since the 1970s has fueled an interest in understanding nonlinear optics in single-mode fibers (SMFs), as nonlinear impairments are now understood to be a critical feature that limits the transmission capacity of optical networks\cite{Mitra2001,Essiambre2010}. In the face of ever-increasing bandwidth demands, spatial division multiplexing (SDM) has emerged as a new frontier to improve network capacity\cite{Richardson2013}.  Although multimode fibers pre-date single-mode fiber, the nonlinear effects in multimode fibers have received comparatively little attention, because until recently MMFs had been primarily relegated to short-distance, low-power links.  As few-mode fibers and multimode fibers become more prevalent in longer-distance networks, and with the advent of techniques for spatially-multiplexed optical amplification\cite{Bai2012}, nonlinear effects are expected to play an increasingly important role\cite{Essiambre2013}.

Multimode nonlinear optics is inherently more complex than single-mode nonlinear optics because of the spatial degree of freedom.  Commercially available MMFs support a large number (\textasciitilde 100s) of transverse spatial modes, which give rise to a variety and number of intramodal and intermodal nonlinear interactions.  A plethora of spatiotemporal nonlinear phenomena have been uncovered in recent experiments, including Kerr-induced beam cleanup in graded-index MMFs \cite{Liu2016,Krupa2017}, multimode solitons \cite{Renninger2013}, geometric parametric instability \cite{Krupa2016}, multi-octave spanning supercontinuum generation \cite{Eftekhar2017} and spatiotemporal modulation instability \cite{Wright2016}. Such phenomena are of great interest not only from a fundamental science perspective, but also in practical applications ranging from high-power beam delivery and high-power fiber lasers to supercontinuum light sources and optical metrology \cite{Picozzi2015}.

Here, we seek to address two key shortcomings in the way multimode nonlinearity is measured and modeled, respectively. The first relates to experimental measurement techniques. Traditional measurement techniques, including spectral and temporal measurements of the entire beam and spatial imaging using CCD/CMOS cameras, all average over two out of the three measurement axes (space, time and spectrum). In the most common type of measurements (spectral measurements coupled with spatial imaging of the output), both the spectrum analzyer as well as the CCD/CMOS camera average over many pulses. As a result, many interesting dynamics that happen within one pulse duration are missed, and the spatiotemporal nature of multimode nonlinearity is not captured. We address this problem by introducing a method for measuring the output MMFs and FMFs in both space and time. Specifically, we raster scan a near-field scanning electron microscope (NSOM) tip in the near-field of the MMF/FMF output end-face, while collecting a time trace at each spatial location. By stitching together the measured spatially-resolved time traces, we demonstrate the temporal evolution of the instantaneous intensity profile within one optical pulse, with a temporal resolution of 40 ps and spatial resolution of 400 nm.

There is growing recognition, backed by experimental measurements, that the nonlinearity in MMFs is a spatiotemporal phenomenon.  For example, Krupa \textit{et al.} report camera images, time-traces and spectra of power-dependent supercontinuum generation in graded-index (GRIN) MMFs \cite{Krupa2016b}, although both the temporal as well as spectral measurements were spatially averaged across the fiber core area\cite{Krupa2016b}.  In \cite{Krupa2018}, Krupa \textit{et al.} used a small-area high-speed photodetector to measure different temporal pulse profile at different positions in the collimated output beam from a graded-index MMF.  Here we report a complete spatiotemporal measurements of the entire output beam, measured directly in the near-field.

The second shortcoming we seek to address here relates to the understanding of the nature of nonlinearity itself. Currently, there exist two complementary models of nonlinear propagation of optical pulses in MMFs: the (3+1)D nonlinear Schr\"{o}dinger partial differential equation for the complex field envelope (also known as the Gross-Pitaevskii equation), and the generalized multimode nonlinear Schr\"{o}dinger equations (GMM-NLSE) \cite{Poletti2008}. The nonlinear wave equation is most efficient for numerically simulating the nonlinear propagation of pulses when a large (\textasciitilde 100s) number of modes are excited \cite{Krupa2016}. The GMM-NLSE on the other hand is best suited for studying and numerically simulating pulse propagation when the number of excited modes is small (\textasciitilde 10s). The GMM-NLSE treats optical nonlinearity, notably different from the case of bulk media, as acting at the modal level. The validity of the latter picture of nonlinearity is of fundamental importance not only in establishing a more complete understanding of multimode nonlinear effects broadly, but also specifically in FMF-based SDM applications where the number of co-propagating modes is small (\textasciitilde 10s). In order to study this problem, we choose a step-index few-mode-fiber, and we further restrict the already small number of allowed modes in FMFs by etching a phase mask directly on the FMF input end-face by means of focused ion-beam (FIB) milling. The phase mask restricts the number of excited modes to the smallest non-trivial number possible: two. We measure the output of this system in space and time simultaneously and compare our results with the predictions of the GMM-NLSE and its modal treatment of nonlinearity.

\section{Selective Mode Excitation}
\label{sel_mode_exc}

Under the weakly-guiding approximation, a linearly-polarized optical signal with carrier frequency $\omega$ traveling in a few-mode optical fiber can be represented as
\begin{equation}\label{eq:1}
    \vvv{E}(r,\phi,z,t) = \xhatv \sum\limits_p A_p(z,t)\psi_p(r,\phi)e^{i(\beta_pz-\omega t)}
\end{equation}
%
where $\psi_p(r,\phi)$ represents an LP mode of the fiber, with corresponding propagation constant $\beta_p$, and $A_p(z,t)$ is the slowly-varying complex envelope.  To simplify the notation, here we contract the azimuthal and radial indices $(l,m)$ into a single index $p$ that enumerates the LP modes.

For simplicity, we consider here the excitation and interaction among the radially symmetric modes (LP$_{0m}$ modes).  For a step-index fiber with core index $n_1$ and cladding index $n_2$, the radially symmetric modes are given by
\begin{equation}\label{eq:3}
  \psi_p(r) = N_p
    \begin{cases}
      J_0\left( U_p r/a \right), & r \le a \\
      K_0\left( W_p r/a \right),  & r > a
    \end{cases},\quad U_p \equiv a\sqrt{\dfrac{n_1^2\omega^2}{c^2} - \beta_p^2},\quad W_p \equiv a\sqrt{\beta_p^2 - \dfrac{n_2^2\omega^2}{c^2}}
\end{equation}
The propagation constants $\beta_p$ and related values $U_p$ and $W_p$ are determined by requiring that $\psi_p(r)$ and $\psi_p'(r)$ be continuous at the boundary $r=a$.  The normalization constant $N_p$ is chosen so that $|A_p(z,t)|^2$ represents the instantaneous power carried the $p$-th mode.

The step-index few-mode fiber considered in this work has a core diameter $2a = 20$ $\mu$m and numerical aperture of 0.14.  At our laser wavelength $\lambda=1064$ nm, this fiber supports 17 LP modes, of which three are radially symmetric: $LP_{01}$, $LP_{02}$ and $LP_{03}$, which we label $p=$ 1, 2 and 3.

If the input face of the optical fiber is illuminated by a symmetrical, focused, linearly polarized, Gaussian optical beam described by $\Phi(r) = \exp(-r^2/w^2)$, then a superposition of the radially symmetric modes will be excited, and the relative portion of power coupled into each of these modes is:
\begin{equation}\label{eq:4}
  \eta_p \equiv \frac{|A_p(z=0,t)|^2}{P_0(t)} = \dfrac{\left|\iint \Phi(r) \psi_p(r) dA\right|^2}{\left|\iint \Phi(r) dA\right|^2\left|\iint \psi_p(r) dA\right|^2}
\end{equation}
where $P_0(t)$ represents the total power of the incident Gaussian beam.

\begin{figure}[hbt]
\centering\includegraphics{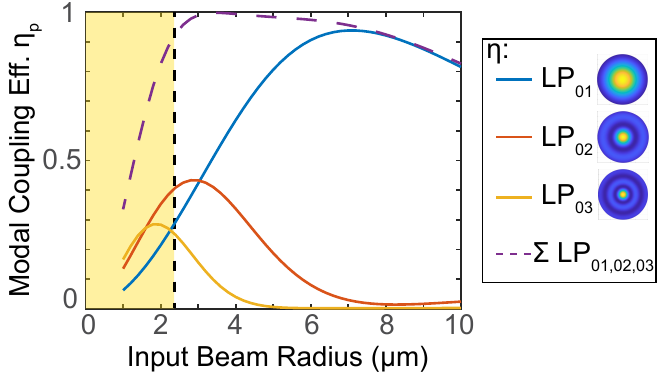}
\caption{Modal coupling efficiencies vs input beam radius. For beam radii to the left of the dashed vertical line, the overall coupling efficiency of light into the FMF falls of quickly. To maintain good overall coupling efficiency, input beam radius must be larger than 2.3 $\mu$m.}
\label{fig:1}
\end{figure}

In Figure \ref{fig:1}, we plot the numerically calculated modal coupling efficiencies $\eta_p$ for the three radially symmetric modes as a function of input beam radius $w$, along with the total coupled power efficiency, $\eta_1+\eta_2+\eta_3$.  The total coupled optical power decreases well below unity for input excitations smaller than $w = 2$ $\mu$m, because in this regime the focused Gaussian beam exceeds the numerical aperture of the fiber.  From Fig.~\ref{fig:1}, one sees that with a simple Gaussian beam is not possible to excite one higher order mode exclusively, and more importantly it is not possible to selectively excite a combination of two modes with comparable powers without also launching significant power in the third mode.  To best isolate and study the nonlinear interaction between propagating modes, we seek a method for selectively, efficiently, and exclusively exciting a pair of modes -- which is impossible with a simple Gaussian input beam.

Selective excitation of spatial modes has been previously achieved using spatial light modulators (SLMs) \cite{Rishoj2019,Zhu2018} together with projection optics. However, spatial modulators can be bulky, difficult to align, and are prone to damage under high fluence illumination that is required to observe nonlinear optical effects. Selective excitation of orbital angular momentum modes has also been achieved using forked diffraction gratings patterned directly on the fiber end-face \cite{Xie2018}, but the diffraction efficiency can limit significantly limit the coupled optical power, which again hinders the observation of nonlinear optical effects.  Here, we employ a new method of directly imparting a binary phase pattern onto the incident Gaussian beam by modifying the fiber end-face.  Our method is inspired by thin-film deposition reported by Chen et al. \cite{Chen1988}, but instead of depositing and patterning thin films onto the input end-face, we directly etch a phase mask onto the input end-face of the FMF using focused ion beam (FIB) milling.

\begin{figure}[hbt]
\centering\includegraphics{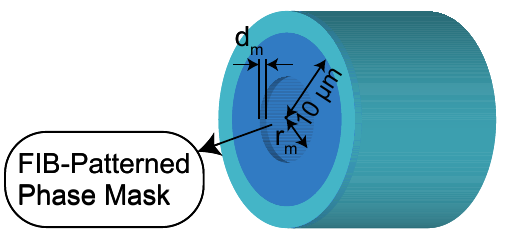}
\caption{Graphical illustration of an FIB-patterned phase mask on the input end-face of the FMF. The highlighted region at the center of the core represents the region where SiO$_2$ was removed in the milling process. (Note: The cladding diameter is not to scale.)}
\label{fig:2}
\end{figure}

Although focused-ion beam milling permits nearly arbitrary spatial structures, we have found that a simple binary radial pattern provides sufficient degrees of freedom to selectively and efficiently excite two radial modes.  Prior to fabrication, the FMF end-face is coated with a 100 nm layer of Au:Pd alloy in order to make the sample conducting in order to help mitigate charging effects during the milling process.  An accelerated beam of Ga$^+$ ions is focused onto the FMF input end-face to a spot size of about 90 nm.  The ion beam is raster-scanned to remove a centered disc pattern of radius $r_m$ and depth $d_m$ as shown in Figure \ref{fig:2}.  Because the core/cladding boundary is not discernable on the endface through electron microscopy, the focused ion beam write-pattern was aligned to the fiber outer diameter.  The removal of SiO$_2$ in the disc region imparts a phase difference $\Theta(r)$ to the near-field coupled light, thereby creating a spatial phase mask described by
\begin{equation}\label{eq:5}
  \Theta(r) = \begin{cases}
    (n_1-1)\dfrac{\omega}{c}d_m,& r \le r_m\\
    0,& r > r_m
  \end{cases}
\end{equation}

\begin{figure}[hbt]
\centering\includegraphics{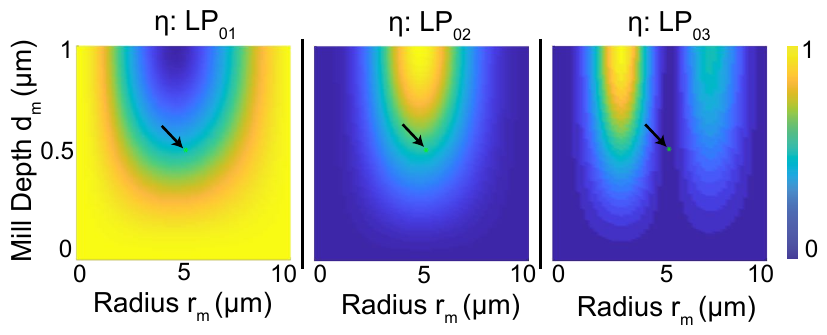}
\caption{Numerical calculation of modal coupling efficiencies as a function of phase mask radius ($r_m$) and depth ($d_m$). The chosen operating point marked by * in green is ($r_m^*$,$d_m^*$)=(5.28 $\mu$m, 0.53 $\mu$m).}
\label{fig:3}
\end{figure}

\begin{figure}[hbt]
\centering\includegraphics{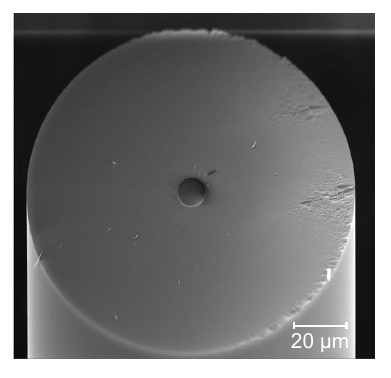}
\caption{Scanning electron micrograph (SEM) of FMF input end-face after FIB milling process.  The darker disc at the center indicates the area where milling was performed.}
\label{fig:4}
\end{figure}

In the presence of a phase mask described by the function $\Theta(r)$, the launched modal amplitudes are re-calculated to be:
\begin{equation}\label{eq:4}
  \eta_p = \dfrac{\left|\iint \Phi(r) e^{i\Theta(r)} \psi_p(r) dA\right|^2}{\left|\iint \Phi(r) dA\right|^2\left|\iint \psi_p(r) dA\right|^2}
\end{equation}
where, as before $\Phi(r) = \exp(-r^2/w^2)$.  Figure \ref{fig:3} shows the calculated coupling efficiency for the three radial modes, as a function of the two degrees of freedom ($d_m$, $r_m$), for a fixed input beam radius of $w = 8.4$ $\mu$m. The radius of the mask is varied from 0 to 10 $\mu$m, while the depth of mask is varied from 0 to 1 $\mu$m (approximately one wavelength). Regions of interest on this color map include those that have negligible power in one mode and comparable power in the other two. The chosen operating point is marked by * (in green) in Figure \ref{fig:3}, at which the $LP_{03}$ color map shows very low coupling efficiency, while $LP_{01}$ and $LP_{02}$ color maps show comparable efficiencies. The calculated modal coupling efficiencies at this point are $\eta_1=0.47$, $\eta_2=0.31$ and $\eta_3<0.01$ for $LP_{01}$, $LP_{02}$ and $LP_{03}$ modes respectively. The final result of this FIB milling process is shown in the scanning electron microscopy (SEM) image of the FMF input end-face shown in Fig.~\ref{fig:4}.

\section{Nonlinear Optics in FMFs: Theory and Modeling}
\label{modeling}

In single-mode fibers, pulse propagation in the presence of the optical Kerr nonlinearity is described by the nonlinear Schr\"{o}dinger equation.  In multimode and few-mode fibers, the single equation must be replaced by the generalized multimode nonlinear Schr\"{o}dinger equation -- a set of coupled partial differential equations that govern the evolution and mixing of the mode amplitudes\cite{Poletti2008}.  The equation governing the $p$-th mode amplitude is:
\begin{equation}\label{eq:7}
  \frac{\partial A_p}{\partial z} =
    \beta_p'\frac{\partial A_p}{\partial t}
    - i\frac{\beta_p''}{2}\frac{\partial^2A_p}{\partial t^2}
    + i\sum_{l,m,n}\gamma_{lmnp}A_lA_mA^*_ne^{i\Delta \beta_{lmnp}z}
\end{equation}
where $\beta_p'$ and $\beta_p''$ are the first and second-order Taylor series coefficients of the propagation constant $\beta_p(\omega)$ about the optical carrier frequency $\omega$, which relate to the group velocity and chromatic dispersion, respectively.  The phase mismatch is given by:
\begin{equation}\label{eq:8}
  \Delta\beta_{lmnp} \equiv \beta_l+\beta_m-\beta_n-\beta_p
\end{equation}
and the nonlinear coefficient governing the mixing between modes is
\begin{equation}\label{eq:9}
  \gamma_{lmnp} \equiv \frac{n_2\omega}{cA_{lmnp}}
\end{equation}
where $n_2$ is the nonlinear refractive index, and $A_{lmnp}$ is an effective area,
\begin{equation}\label{eq:10}
  A_{lmnp} \equiv \frac{\sqrt{\int|\psi_l|^2dA\int|\psi_m|^2dA\int|\psi_n|^2dA\int|\psi_p|^2dA}}{\int \psi_l^*\psi_m\psi_n^*\psi_pdA}
\end{equation}
Earlier treatments \cite{Poletti2008} of this problem assumed a common $z$-dependence of $e^{i\beta_0z}$ for all modes in \eqref{eq:1}, which eliminates the phase mismatch $\Delta\beta_{lmnp}$, but instead introduces an additional term $i(\beta_p-\beta_0)A_p$ on the right-hand-side of \eqref{eq:7}.  The equivalent formulation presented here is more convenient and easier to numerically integrate for longer optical pulses.

For the few-mode fiber considered here, there are 17 modes per polarization state at $\lambda = 1064$ nm, and therefore \eqref{eq:7} represents a set of 17 coupled nonlinear equations, with up to $17^4 = 83,521$ nonlinear coupling coefficients $\gamma_{lmnp}$ and phase mismatch terms $\Delta\beta_{lmnp}$.  Fortunately, many of the nonlinear terms vanish because of the symmetries of the interacting modes.  Of those remaining, many have a non-zero phase mismatch, and therefore average to zero over the length of the fiber considered here \cite{Horak2012}.  For the pulse durations and fiber lengths considered here, the chromatic dispersion can be safely ignored ($\beta_p'' = 0$), and we can further ignore the differential group delay between the interacting modes, i.e., we assume that $\beta_p'$ are all equal.  Finally, as long as the coherent mixing terms (i.e., those with nonzero $\Delta\beta$) are ignored, the power in each mode remains constant, and hence any mode that is not excited at $z=0$ will remain absent as the field propagates.  Therefore, if only two LP modes are exclusively excited at $z=0$, the generalized nonlinear Schr\"odinger equation simplifies to a pair of coupled equations for these two modes:
\begin{align}
  \label{eq:11}
  \frac{\partial A_1}{\partial z} - \beta'\frac{\partial A_1}{\partial t} &= i \left(\gamma_{1111}|A_1|^2 + 2\gamma_{1122}|A_2|^2\right)A_1 \\
  \label{eq:12}
  \frac{\partial A_2}{\partial z} - \beta'\frac{\partial A_2}{\partial t} &= i \left(\gamma_{2222}|A_1|^2 + 2\gamma_{1122}|A_1|^2\right)A_2
\end{align}
where the coefficients $\gamma_{1111}$ and $\gamma_{2222}$ describe self-phase modulation and $\gamma_{1122}$ describes the process of cross-phase modulation between the two excited modes.  These equations can be directly integrated, to yield
\begin{align}
  \label{eq:13}
  A_1(L,\tau) &= A_1(0,\tau) \exp\left[i(\gamma_{1111}|A_1|^2 +2\gamma_{1122}|A_2|^2)L\right]\\
  \label{eq:14}
  A_2(L,\tau) &= A_2(0,\tau) \exp\left[i(\gamma_{2222}|A_2|^2 +2\gamma_{1122}|A_1|^2)L\right]
\end{align}
where $\tau\equiv t - \beta'L$ is the retarded time.  In general, the two modes will incur different nonlinear phase shifts as the wave propagates, and the relative phase between the two modes emerging from the fiber will be power-dependent.  The spatial intensity emerging at the end of the fiber is given by
\begin{equation}
    \label{eq:15}
	I(r,z=L,t) \propto \left|\psi_1(r)A_1(L,\tau)e^{i\beta_1L} + \psi_2(r)A_2(L,\tau)e^{i\beta_2L}\right|^2
\end{equation}
Because the transverse modes differ, the degree of interference between them will be spatially dependent and because the pulse shape is time-dependent, it produces a time-varying nonlinear phase shift between the modes.  These effects combine to produce a spatiotemporal nonlinear pattern at the output face.  If the input pulse is Gaussian in time, it will split into two spatial modes, which each acquire different nonlinear chirps, as the pulse intensity rises and then falls.  If the peak nonlinear phase difference approaches or exceeds $\pi$, the local intensity emerging from the fiber will exhibit a pattern of temporal interference fringes associated with the turn-on and turn-off of the pulse.

These broad predictions of the analytical model (based on certain simplifying assumptions) form a starting point with which to compare the results of our numerical simulations and experimental results. In the following sections, we present our experimental setup, measurements, and compare them numerical Split-Step Fourier Method (SSFM) simulations of \eqref{eq:7}.

\section{Experiment}

The experimental setup consists of a YAG microchip laser ($\lambda_0=1064$ nm) that produces 720 ps pulses at a 1 kHz repetition rate. The laser pulses have an energy of about 135 $\mu$J, and the energy of the pulses going into the FMF is controlled by a series combination of a half-wave plate (HWP) followed by a polarizing beam splitter (PBS) such that the input peak power is 15 kW. Using a plano-convex lens of focal length f = 25.4 mm, the laser beam is focused down to a spot with radius 8.4 $\mu$m on the patterned input end-face of a 20 $\mu$m step-index FMF of numerical aperture $NA=0.14$ and length L=1.24 m. The patterning has been done in order to selectively excite the $LP_{01}$ and $LP_{02}$ modes with comparable amplitudes, as described in Section \ref{sel_mode_exc}.

At the output end-face of the FMF, we employ a near-field scanning optical microscope (NSOM) tip that is brought in close proximity ($\ll$ 1 $\mu$m) to the FMF end-face. The NSOM tip has an aperture of 250 $\pm$ 50 nm, and tapers into a single-mode fiber segment that is connected to a 10 GHz photo-receiver and recorded using a real-time oscilloscope. The resolution of the temporal measurements is 40 ps. The NSOM tip is scanned across the output end-face of the FMF using a piezo-controlled translation stage. By using this setup, we record the temporal output along a 20 $\mu$m $\times$ 20 $\mu$m grid of pixels on the FMF output end-face at a resolution of a 400 nm. We then reconstruct a temporal evolution of the 2-D intensity profile exiting the FMF end-face. 


Because this technique involves synchronous measurements of the time-domain waveform at multiple pixels on the FMF output end-face, it is important to eliminate pulse-to-pulse fluctuations and timing jitter. To minimize any timing jitter associated with the pulse detection, we reflected 1\% of the incident pulse to a second high-speed PIN photodiode, which was used to confirm the pulse energy stability, and also as a clean trigger signal for the oscilloscope.

\begin{figure}[hbt]
	\includegraphics{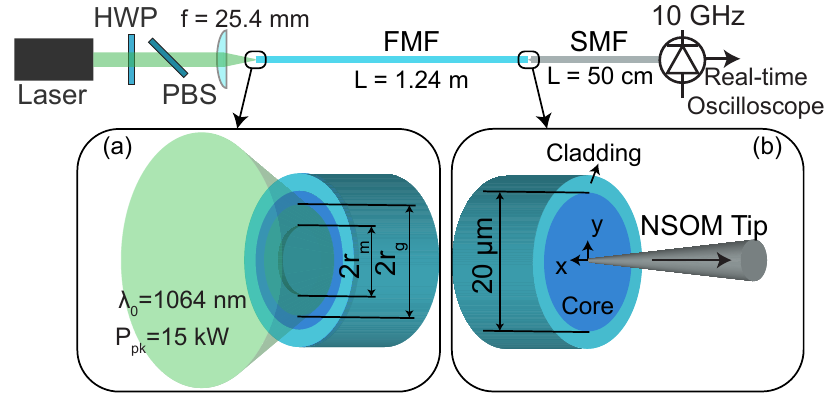}
	\centering
	\caption{Experimental schematic for performing spatiotemporal measurements of nonlinear interactions between selectively-excited spatial modes of a FMF. (a) A Gaussian beam of radius r$_g$ = 8.4 $\mu m$ is incident on the patterned FMF input end-face; mask radius r$_m$ = 5.28 $\mu m$. (b) Raster-scanned NSOM fiber tip for recording spatiotemporal measurements at FMF output end-face; the separation between the FMF end-face and NSOM tip is $\ll$1 $\mu$m.}
	\label{fig:5}
\end{figure}

\section{Results and Discussion}

Using the scanning NSOM-tip method described above, the time-domain output is recorded at different spatial locations on the FMF output end-face. Figure \ref{fig:5}(a) shows, for reference, the radial profiles of the two selectively excited modes: $LP_{01}$ and $LP_{02}$. Figures \ref{fig:5}(b) through \ref{fig:5}(d) show the time-domain output recorded at 3 selected spatial locations: $r =$ 0 (on-axis), $r =$ 4.4 $\mu$m, and $r =$ 7.2 $\mu$m respectively. As discussed in Section \ref{modeling}, we observe interference fringes in the time-domain arising from the overlap, in time and space, of two modes that have acquired different nonlinear chirps. Further, at the three selected values of $r$, the two modes have different magnitudes and phases, leading to a different time-domain pattern at each value of $r$.

\begin{figure}[hbt]
	\includegraphics{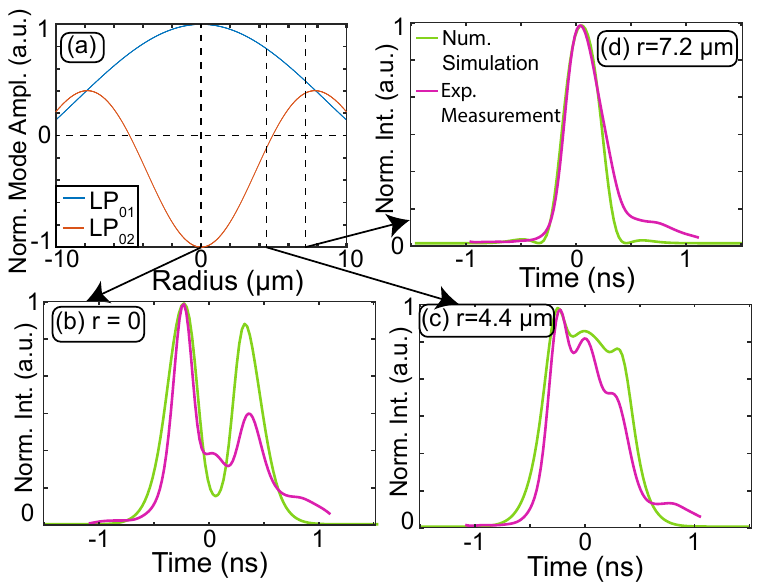}
	\centering
	\caption{Spatially-resolved Temporal Measurements: (a) Radial amplitude profiles of $LP_{01}$ and $LP_{02}$ modes; (b), (c) and (d) Output temporal measurements at $r=0$ (on-axis), $r = 4.4$ $\mu$m and $r = 7.2$ $\mu$m.}
	\label{fig:6}
\end{figure}

\begin{figure}[hbt]
	\includegraphics{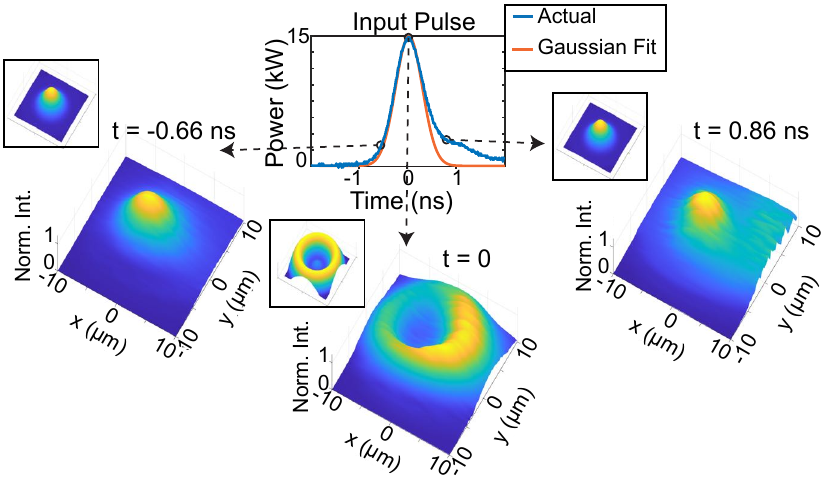}
	\centering
	\caption{Spatiotemporal reconstruction of output pulse at FMF output end-face: Output 2D spatial intensity profile (reconstructed from experimental data) at 3 time instances within the pulse: $t = -0.66$ ns, $t = 0$ and $t = 0.86$ ns. (See Supplementary Video 1). Numerical simulation results are shown in inset boxes. (See Supplementary Video 2).}
	\label{fig:7}
\end{figure}

The temporal data is then used to reconstruct the 2D spatial intensity pattern at the FMF output for time-instances within one pulse duration. By stitching together all of the time traces into a mosaic of time-varying pixels, we construct what one would see at the output with a piconsecond-scale ultrafast video camera. Figure \ref{fig:6} shows the output of such a reconstruction, compared with the input Gaussian pulse for reference.

In order to make better understand the temporal and spatial features observed in the results shown in Figures \ref{fig:6} and \ref{fig:7}, we note that in the absence of any nonlinearity, the net output spatial profile is determined by the linear phase difference acquired by the two modes during propagation, which in turn is determined by modal propagation constants and the exact length of the FMF. In this case, the length of the fiber is such the two modes interfere in the absence of any nonlinearity to produce a net output spatial profile that is Gaussian-like, with an on-axis maximum. This is seen at t=-0.66 ns and t=0.86 ns in Figure \ref{fig:7}, where the instantaneous power of the pulse (and therefore the nonlinearity) is low.

As the pulse rises to its peak, nonlinearity can no longer be neglected, as the two modes acquire a time-dependent nonlinear phase difference. Using the expressions for the time-dependent nonlinear phases acquired by the two modes from Section \ref{modeling}, $\Gamma_1(t)$ and $\Gamma_2(t)$, the peak nonlinear phase difference the two modes acquire at the pulse peak (t=0) is calculated to be roughly equal to one $\pi$ for our experimental parameters. This $\pi$ phase difference has two key manifestations. First, as shown in Figure \ref{fig:7}, at t=0, the on-axis maximum is converted to an on-axis minimum, thereby converting a Gaussian-like beam to an annulus-shaped beam. Second, at the on-axis point, the \textit{same} destructive interference of the modes manifests in the form of a local minimum at t=0, resulting in the formation of the temporal fringes as shown in Figure \ref{fig:6}(b).

\subsection{Validity of modal treatment of nonlinearity}

Having resolved the output in space as well as time, we are now in a position to address the validity of the modal treatment of nonlinearity for our experimental conditions. We do this in two different ways: we first directly compare our experimental results with the predictions of numerical simulations of GMM-NLSE \eqref{eq:7}.

\subsubsection{Theory (GMM-NLSE) vs Experiment}

The Bessel functions describing the modal profiles of the $LP_{01}$ and $LP_{02}$ modes have opposite phases for $r<4.92$ $\mu$m, where the $LP_{02}$ mode has a null. For $r>4.92$ $\mu$m, they are in phase. This is shown in Figure \ref{fig:6}(a). As discussed above, the net phase difference (including contributions from the Bessel functions, from linear as well as nonlinear propagation) between the two modes at the on-axis location is such that the interference of the modes causes a local minimum in the normalized temporal intensity profile at t=0. However, for $r>4.92$ $\mu$m, the $LP_{02}$ Bessel function flips in phase, adding an additional $\pi$ phase difference between the modes. At the $r=7.2$ $\mu$m spatial location, we indeed observe a local \textit{maximum} at t=0, as opposed to a local minimum as at $r=0$. This is shown in Figure \ref{fig:6}(d). This suggests that the nonlinear component of the phase difference remains the same at both the spatial locations. This less obvious feature of the spatially-resolved temporal data confirms the validity of the modal treatment of nonlinearity. Further, our experimental results are consistent with the numerical simulations of the GMM-NLSE \eqref{eq:7}, as shown in Figures \ref{fig:6} and \ref{fig:7}. This further suggests that the modal treatment of nonlinearity offered by the GMM-NLSE is valid.

One final note we make about the experimental and numerical results of the output temporal waveforms in Figure \ref{fig:6} is the following: the asymmetry about t=0 results from the slight difference in group delays over the length of the FMF, which we have neglected in the simplified analytical model presented in Section \ref{sel_mode_exc}, but included in our numerical simulations of the GMM-NLSE. The imperfect shape of the laser pulse itself also causes asymmetric features, and was not accounted for in our numerical simulations.

\subsubsection{Thermal Measurements}

Our GMM-NLSE-based model explains our findings as follows: as the pulse rises towards its peak, the instantaneous phase difference between the two modes varies with time as they acquire different instantaneous nonlinear phase shifts via SPM and XPM. As a result, the instantaneous spatial pattern formed by their interference also varies with time.

In order to further verify that the spatiotemporal features that we observe are a result of interference of the modes, we performed the following test. We found a different mechanism -- different from optical nonlinearity -- to vary the relative phase between the two modes at the output end-face of the FMF, and compared the output intensity patterns from this test to what our spatiotemporal measurements show. We achieved this by heating a section of the fiber on a hot plate. The rise in temperature causes two effects: i) thermal expansion of the length of the fiber, and ii) change in refractive index of the fiber core, which causes a change in modal propagation constants. However, the thermal effect on the modal propagation constants would be the same for both the modes. Since the relevant quantity for interference of the two modes is the \textit{difference} in their propagation constants, the second of the effects listed above can be neglected for the purposes of these measurements.

\begin{figure}[hbt]
	\includegraphics{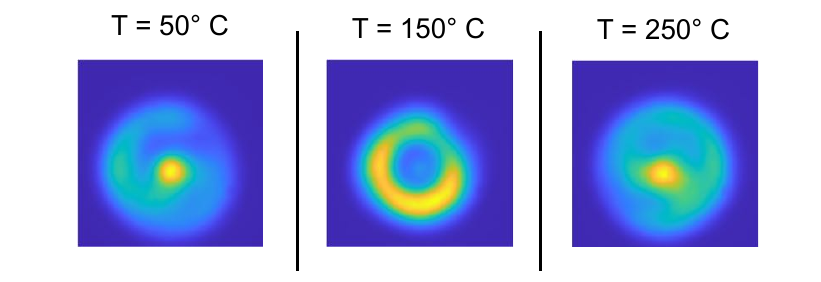}
	\centering
	\caption{FMF output recorded on a CMOS camera at low input power at 3 different temperatures. As the temperature of a 20 cm long FMF section is increased, the length of the core increases on the micron-scale due to thermal expansion, leading to a slightly different modal overlap at each temperature. As the temperature is swept from 50$^0$ C to 150$^0$ C, the output intensity profile switches between a Gaussian-like shape and an annulus, just as it did within one pulse duration in the presence of nonlinearity}
	\label{fig:8}
\end{figure}

In the absence of any optical nonlinearity, the difference in linear phases acquired by the two modes at the output is given by $\theta_{1,2}=(\beta^{(1)}_0-\beta^{(2)}_0)$L. In the presence of heat, L changes slightly, resulting in a change in the phase difference with which the two modes overlap at the output, thereby resulting in a temperature-dependence of the interference pattern. This change in the output intensity pattern is easily observed on a CMOS/CCD camera. Figure \ref{fig:8} shows the FMF output recorded on a CMOS camera at 3 different temperatures (corresponding to 3 different values of relative phase between the two modes) at low input power. As one can see, the output intensity profile varies from a Gaussian-like shape to an annulus and back; exactly as in our spatiotemporal measurements of nonlinearity. (Note that when the fiber is stretched by heating, it introduces a relative phase difference between the two modes that remains constant through the duration of a pulse. Additionally, as the input power is kept low for these measurements to keep out any nonlinear effects, there is no time-dependent phase introduced. As a result, it is sufficient to use a CMOS camera, which averages over many pulses, to capture the output intensity profile.)

The agreement between the predictions of our simplified analytical model, numerical simulations of the complete GMM-NLSE \eqref{eq:7}, spatiotemporal measurements, as well as thermal measurements indicates the validity of the modal picture of nonlinearity.

\section{Conclusion}

In order to probe the nonlinear interaction between two individual spatial modes, it is desirable to preferentially excite only the modes of interest. We have achieved this using a novel implementation of a phase mask, that involves etching the mask directly onto the fiber input end-face by means of FIB milling. While hard-writing a mask onto the input end-face has the disadvantage of being less flexible as compared to an SLM setup, it has some key advantages such as compactness and ease of integration into chip-scale photonic circuits. Such a mask is also not prone to damage under the influence of high laser power, and is a power efficient way of exciting a desired mode combination.

Having excited the desired combination of modes, we measured the output in both space and time. Our spatiotemporal measurement technique that employs a raster-scanned NSOM-tip brings out the rich spatiotemporal nonlinear dynamics that happen within the duration of a single pulse that are not possible to observe using traditional CCD/CMOS cameras and optical spectrum analyzers. For the case of two $LP_{0m}$ modes excited in a step-index FMF, our measurements demonstrate the existence of interference fringes in the time-domain output, as predicted by the GMM-NLSE. Further, upon using the raster-scanned measurements to reconstruct the temporal evolution of the instantaneous intensity profile at the FMF output end-face, we see that the instantaneous intensity profile transforms from a Gaussian-like shape to an annulus, and back, as the pulse rises, peaks and falls. These form the first complete spatiotemporal measurements of multimode nonlinearity to our knowledge.

In order to confirm that the phenomenon that we observe is a result of a time-varying relative phase difference between two overlapping spatial modes, we varied the relative phase difference by using a mechanism different from optical nonlinearity (i.e. by heating a section of the fiber) and observed the change in output intensity pattern. The change in output intensity pattern with temperature was seen to be consistent with spatial interference of two modes that have acquired different time-dependent nonlinear phases. Further, the match in analytical and numerical models, and spatiotemporal and thermal measurements demonstrate the validity of the modal picture of nonlinearity.

The methods presented here could find applications in further study of spatiotemporal nonlinear phenomena reported previously, such as Kerr-induced beam cleanup and supercontiuum generation in GRIN MMFs, to help shed more light on the mechanisms behind them. Future directions of work could also include resolving the output not just in space and time but also in polarization. Spatiotemporal measurements of the output could also be very useful in the development of multimode spatiotemporally mode-locked lasers, where both the output mode quality and the output pulse shape are of interest.

The effects of optical nonlinearity in MMFs and FMFs are fundamentally spatiotemporal in nature. In order to best understand the physics of these systems and the nonlinear dynamics that arise in them, a full (2+1)D diagnostic that can measure in both space and time (or frequency) is required \cite{Krupa2019}. The near-field measurement technique that we presented here serves as a promising tool with which to better understand nonlinear optics in MMFs and FMFs.

\section*{Acknowledgments}
We acknowledge the support of the Maryland NanoCenter and its AIMLab.

\bibliography{Dacha-Optica-2020}

\end{document}